\documentclass[prl,twocolumn, superscriptaddress,showpacs]{revtex4-1}%
\usepackage{amsfonts}
\usepackage{amsmath}
\usepackage{amssymb}
\usepackage{graphicx}%
\usepackage[english]{babel}
\newcommand{\beq}{\begin{equation}}
\newcommand{\eeq}{\end{equation}}
\newcommand{\nn}{\nonumber}
\newcommand{\ket}[1]{|#1\rangle}
\newcommand{\bra}[1]{\langle #1|}
\newcommand{\up}{\uparrow}
\newcommand{\dn}{\downarrow}

\usepackage{babel}

\begin{document}

\title{Optimal rectification by strongly coupled spins}

\author{T. Werlang
}\email{thiago\_werlang@fisica.ufmt.br}

\author{M. A. Marchiori
}
\author{M. F. Cornelio
}

\author{D. Valente
}


\affiliation{
Instituto de F\'isica, Universidade Federal de Mato Grosso, Cuiab\'a MT, Brazil}

\begin{abstract}
We study heat transport in a pair of strongly coupled spins. In particular, we present a condition for optimal rectification, i.e., flow of heat in one direction and complete isolation in the opposite direction.
We show that the strong-coupling formalism is necessary for correctly describing heat flow in a wide range of parameters, including moderate to low couplings. We present a situation in which the strong-coupling formalism predicts optimal rectification whereas the phenomenological approach predicts no heat flow in any direction, for the same parameter values.
\end{abstract}
\maketitle

%
Manipulation of individual quantum systems represents a breakthrough in the physical sciences \cite{breakthrough}.
It has been successfully achieved with single atoms \cite{haroche,rempe}, ions \cite{dwineland} or molecules \cite{sandoghdar}, and more recently with artificial atoms, like quantum dots \cite{jmgerard,pascale,vuckovic1,finley,reviewphotonics}, or superconducting qubits \cite{devoret,wallraff}.
It opens perspectives in quantum information processing, motivating studies on light-matter interaction at the single-photon level \cite{kojima,Chang,AlexiaVuckovic2,DV,vuckovic3}.
In analogy to modern electronic circuits, quantum devices have been proposed such as photon diodes \cite{dudu,AlgumCitadoEmDudu} and photon transistors \cite{hwang,astafiev}.
Diodes are current rectifiers. 
An optimal rectifier is able to conduct current in one sense and isolate it in the opposite sense.

All such realistic quantum systems are, of course, open.
Natural atoms interact with electromagnetic environments \cite{purcell}. 
Artificial atoms also interact with their solid-state environment. 
There is the need to understand, at the single-quantum level, for instance, the influence of temperature \cite{briegel,jaksch,heatt} and of phonons \cite{huangrhys,PRBdv}, fluctuating charges \cite{robson}, nuclear or electronic spins \cite{pbertet}.
Analogies to diodes and transistors are also extendable to the flow of all such complex excitations \cite{list}.

Manipulation of individual quantum systems also gave birth to engineered interactions between those systems \cite{devoret2}. 
In particular, ultra-strong couplings are achieved, e.g., between a two-level system and a single-mode cavity in circuit QED \cite{ultrastrong}, totally modifying standard quantum optical scenarios \cite{termoscontragirantesJC_TW}.

In this paper, we explore heat transport under the influence of strong coupling between spins. 
We argue that the strong-coupling formalism is necessary even for moderate and low couplings. 
We treat a case where optimal rectification is expected within the strong-coupling description and is completely absent for the standard phenomenological approach.
Optimal rectification is evidenced by the system of two spins coupled via Ising interaction. 
A broad range of experiments is capable of reproducing Ising-type interactions, simulating spins in the strong-coupling regime \cite{RefThiagoFred}.

\begin{figure}
\begin{center}
\includegraphics[width=0.99\linewidth]{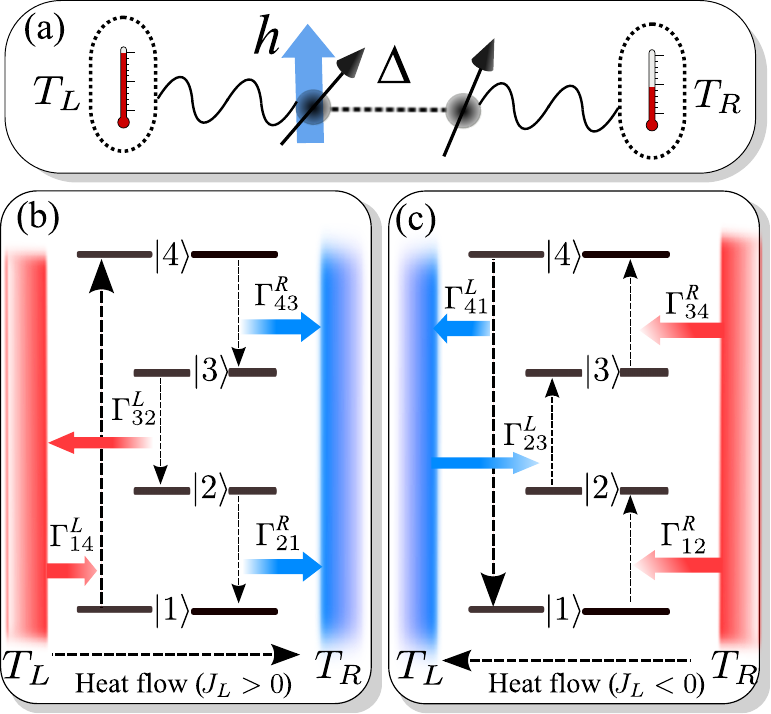}
\caption{
(a) Two spins coupled with strengh $\Delta$. 
Two independent thermal reservoirs are defined at temperatures $T_L$ ($T_R$) for
the bath on the left (right). (b) and (c): case $\Delta < h$. $\ket{1}$ to $\ket{4}$ are the eigenstates of $H_S$ in the corresponding case.
(b) Energy transitions that allow for heat flow from the left hot bath to the right cold bath ($J_L>0$, as defined in the text). 
$\Gamma_{ij}^{L(R)}$ is the net rate of transition from state $\ket{i}$ to $\ket{j}$, driven by the left (right) bath. The colored thick arrows indicate the flow of heat involved in each step of the cycle.
(c) The same as (b) with reverted temperatures, and correspondingly reverted cycle. Note that in both cases the transition between states $\ket{2}$ and $\ket{3}$ involve heat exchange in the opposite sense with respect to the net heat flow. This is the key ingredient for the establishment of rectification, as explained in the text.
}
\label{fig:scheme}
\end{center}
\end{figure}

{ \it Model.}
The system of interest consists in a pair of interacting spins. We define the coupling constant $\Delta$ between the spins in the $z$-direction. The magnetic field $h$ applied to the spin on the left is also in the $z$-direction. The Hamiltonian of the system is
\beq
H_S = \frac{h}{2}\ \sigma^{L}_z + \frac{\Delta}{2}\ \sigma^{L}_z \sigma^{R}_z.
\eeq
The spin on the left (right) is coupled with a thermal reservoir at a given temperature $T_L$ ($T_R$). 
The system is illustrated in Figure \ref{fig:scheme}(a).
The four eigenstates of $H_S$ are given in terms of the eigenstates of $\sigma^\mathrm{L(R)}_z$, $\ket{\up}$ and $\ket{\dn}$, in decreasing energy order for the case of interest, $\Delta<h$,
$
\ket{4}  = \ket{\up\up}, 
\ket{3}  = \ket{\up \dn}, 
\ket{2}  = \ket{\dn \dn}, 
\ket{1}  = \ket{\dn \up}
$
We define the transition frequencies as 
$
\omega_\mathrm{mn} = \epsilon_m - \epsilon_n,
$
where $\epsilon_k$ is the eigenvalue of $H_S$ for the eigenstate $\ket{k}$. 
In the present case, they read
$
\omega_{41} = h+\Delta, 
\omega_{32} = h-\Delta, 
\omega_{43} = \omega_{21} = \Delta,
\omega_{31} = \omega_{42} = h.
$

%
%
The coupling to the each bath of harmonic oscillators is given by the spin-boson model in the $x$-component, 
\beq
H_{spin-res}^{L(R)} = \sigma^{L(R)}_x \sum_k g_k (a_k^{L(R)} + a_k^{L(R)\dagger}),
\label{Hint}
\eeq
with identical coupling strengths $g_k$. The Hamiltonians of the two free reservoirs are $H_{res,L(R)} =  \sum_k \omega_k a_k^{L(R)\dagger} a_k^{L(R)}$. The model in Eq. (\ref{Hint}) implies that the left (right) bath can only induce transitions on left (right) spin. Therefore, transitions $\ket{4} \leftrightarrow \ket{2}$ and $\ket{3} \leftrightarrow \ket{1}$, which simultaneously flip both spins, are forbidden. The rates at which the remaining transitions occur are computed in the following.

%
%
We derive a master equation to describe the system evolution.
Here comes the crucial step for what follows: strong-coupling formalism ($\Delta\sim h$) is employed \cite{breuer}. 
This means that the Lindbladians are obtained on the basis of the eigenstates of the full system Hamiltonian $H_S$. Consequently, the dissipation mechanism of each spin depends not only on the coupling to its own bath but also on the coupling between the spins themselves. In a phenomenological approach, the Lindbladian derived for a single spin is joined to the master equation that describes the whole chain of spins as if each bath acted in a completely independent manner, disregarding the presence of other coupled spins. This is usually applied for extremely low couplings between the subsystems as compared to the transition frequency of the free subsystem, $\Delta \lll h$. An example is found in typical quantum optical scales, where $\Delta/h \sim 10^{-11} - 10^{-6}$ \cite{breakthrough}. By contrast, the  master equation derived here is valid even in the strong coupling regime. In the Born-Markov approximation with respect to the reservoirs, it reads \cite{breuer}
\beq
\frac{d\rho}{dt} =-i [H_S,\rho] + \mathcal{L}_L[\rho] + \mathcal{L}_R[\rho],
\label{master}
\eeq
in $\hslash = 1$ units, where the Lindblad operators $\mathcal{L}_{L,(R)}[\rho]$ are given by
\begin{eqnarray}
{\cal L}_{L(R)}[\rho] & = & \sum_{\omega>0} \mathcal{J}(\omega) (1+n^{L(R)}_\omega) 
\bigg[ A_{L(R)}(\omega)\rho A_{L(R)}^{\dagger}(\omega)\nn\\
                                & &- \frac{1}{2}\Big\{ \rho,A_{L(R)}^{\dagger}(\omega)  A_{L(R)}(\omega) \Big\} \bigg] \nn \\
                                     & & + \mathcal{J}(\omega) \  n^{L(R)}_\omega
                                     \bigg[A_{L(R)}^\dagger(\omega) \rho A_{L(R)}(\omega)\nn\\
                                    & &  - \frac{1}{2}\Big\{\rho,A_{L(R)}(\omega) A_{L(R)}^{\dagger}(\omega)\Big\} \bigg],
\label{Lindbladians}
\end{eqnarray}
where $\omega=\epsilon_j - \epsilon_i>0$.
The average number of excitations in each reservoir is given by the Bose-Einstein distribution, $n^{L(R)}_\omega = \left[\exp{\frac{\omega}{k_B T_{L(R)}}}-1\right]^{-1}$.
$A_{L(R)}(\omega)=\sum_{\omega=\epsilon_j-\epsilon_i}\left|i\right\rangle\left\langle i\right|\sigma_x^{L(R)}\left|j\right\rangle\left\langle j\right|$ is the Lindblad operator associated with the transition driven by the bath from the left (right), with positive frequency $\omega$. For $\Delta<h$, 
${\cal L}_L[\rho]$ contains two non-vanishing operators, namely, 
$
A_{L}({\omega_{41}})=\ket{1}\bra{4}$ 
and 
$A_{L}(\omega_{32})=\ket{2}\bra{3}$, 
along with their adjoints.
Because $\omega_{43}=\omega_{21}$, the only pair of non-vanishing operators for ${\cal L}_R[\rho]$ is 
$A_{R}(\omega_{43})=\ket{3}\bra{4}+\ket{1}\bra{2}$ and $A_{R}^\dagger(\omega_{43})$.
The baths are chosen to be ohmic, so the spectral functions are linear, $\mathcal{J}(\omega)=\kappa \omega$, where
the constant $\kappa$ is the same for both reservoirs.

%
%
{\it Definition of the heat current.}
Heat flow is characterized by heat current $J_\mathrm{heat}$, defined with the aid of a continuity equation for the average energy going through the system $\langle H_S\rangle$ \cite{briegel,jaksch},
\beq
\frac{\partial}{\partial t} \langle H_S \rangle = -\nabla \cdot J_\mathrm{heat} = - (J_R - J_L).
\label{cont}
\eeq
The LHS of Eq.(\ref{cont}), 
$
\frac{\partial}{\partial t} \langle H_S \rangle = \frac{\partial}{\partial t} \mbox{Tr}\{\rho  H_S\} = \mbox{Tr}\{\dot{\rho}H_S\},
$
is calculated by the use of Eq. (\ref{master}), providing
\beq
\frac{\partial}{\partial t} \langle H_S \rangle = \mbox{Tr}\{\mathcal{L}_L[\rho] H_S\}  + \mbox{Tr}\{\mathcal{L}_R[\rho] H_S\}. 
\label{heatinout}
\eeq
The rate of increase in the average energy of the system is then the sum of the input energy rate $J_{L(R)}^{\mathrm{in}}$ from the left (right) reservoir, 
\beq
J^\mathrm{in}_{L(R)} \equiv \mbox{Tr}\{\mathcal{L}_{L(R)}[\rho] H_S\} = \pm J_{L(R)}.
\label{jin}
\eeq

%
%
{\it Steady-state regime.}
We solve Eq.(\ref{master}) in the steady-state regime, defined as $\dot{\rho}_{ss}=0$, for which $\frac{\partial}{\partial t} \langle H_S \rangle = 0$, $J_L=-J_R$.
In this case, the density matrix is diagonal in the energy eigenstates, $[H_S,\rho_{ss}]=0$.
So Eq. (\ref{master}) reduces to
\begin{eqnarray}
\dot{\rho}_{11} = 0 & = & \Gamma^L_{41} - \Gamma^R_{12} \label{rho11}, \label{rho11} \\
\dot{\rho}_{22} = 0 & = & -\Gamma^L_{23} + \Gamma^R_{12} \label{rho22}, \\
\dot{\rho}_{33} = 0 & = & \Gamma^L_{23} - \Gamma^R_{34} \label{rho33}, \\
\dot{\rho}_{44} = 0 & = & -\Gamma^L_{41} + \Gamma^R_{34} \label{rho44} \label{rho44},
\end{eqnarray}
where
\beq
\Gamma^{L(R)}_{ij} \equiv \kappa\omega_{ij}[(1+n^{L(R)}_{\omega_{ij}})\rho_{ii} - n^{L(R)}_{\omega_{ij}}\rho_{jj}],
\label{effectivedecay}
\eeq
defined for $i>j$.
The first term in the RHS of Eq. (\ref{effectivedecay}) is the decaying rate from state $\ket{i}$ to $\ket{j}$ and the second one is the excitation rate from $\ket{j}$ to $\ket{i}$.
So $\Gamma^{L(R)}_{ij}$ is the net decaying rate from the state $\ket{i}$ to the state $\ket{j}$.
For $j>i$, $\Gamma_{ij}^{L(R)} = -\Gamma^{L(R)}_{ji}$. See Fig.\ref{fig:scheme}(b) and (c).
System (\ref{rho11})-(\ref{rho44}) shows that
$
\Gamma^L_{41} = \Gamma^R_{12} = \Gamma^L_{23} = \Gamma^R_{34} \equiv \Gamma.
$
That is, a single function $\Gamma$ of the parameters $\Delta$, $h$ and $T_{L(R)}$ 
governs the rate that the cycle runs (see \cite{sup} for full expression of $\Gamma$). Fig.\ref{fig:scheme}(b) represents the case where $\Gamma < 0$, and (c), $\Gamma > 0$.

%
%
The steady-state current is computed,
\[
J_L = -\omega_{41} \Gamma^L_{41} + \omega_{32}\Gamma^L_{23} = - 2\ \Delta\ \Gamma,
\]
where we have used that $\omega_{41}=h+\Delta$ and $\omega_{32}=h-\Delta$.
In other words, the left bath absorbs from the left spin the amount $\omega_{41}$ of energy at rate $|\Gamma|$ and delivers $\omega_{32}$ of energy at the same rate, resulting in a net amount of $2\Delta$ of exchanged energy per cycle. $J_L>0$ is obtained when $-\Gamma > 0$, i.e., $-\Gamma = \Gamma^L_{14} > 0$. The last inequality means that energy leaves the left bath and goes to the left spin. Therefore, heat flows from the left to the right reservoir when $J_L>0$ (Fig.\ref{fig:scheme}(b)). Heat current on the right spin is also computed, $J_R = \omega_{21}\Gamma^R_{12}+\omega_{43}\Gamma^R_{34}$. Using the definition of $\Gamma$ and that $\omega_{41}=\omega_{43}+\omega_{32}+\omega_{21}$, we verify that $J_R = (\omega_{21}+\omega_{43})\Gamma=(\omega_{41}-\omega_{32}) \Gamma = \omega_{41}\Gamma^L_{41}-\omega_{32}\Gamma^L_{23} = -J_L$, indeed.

\begin{figure}
\begin{center}
\includegraphics[width=1.00\linewidth]{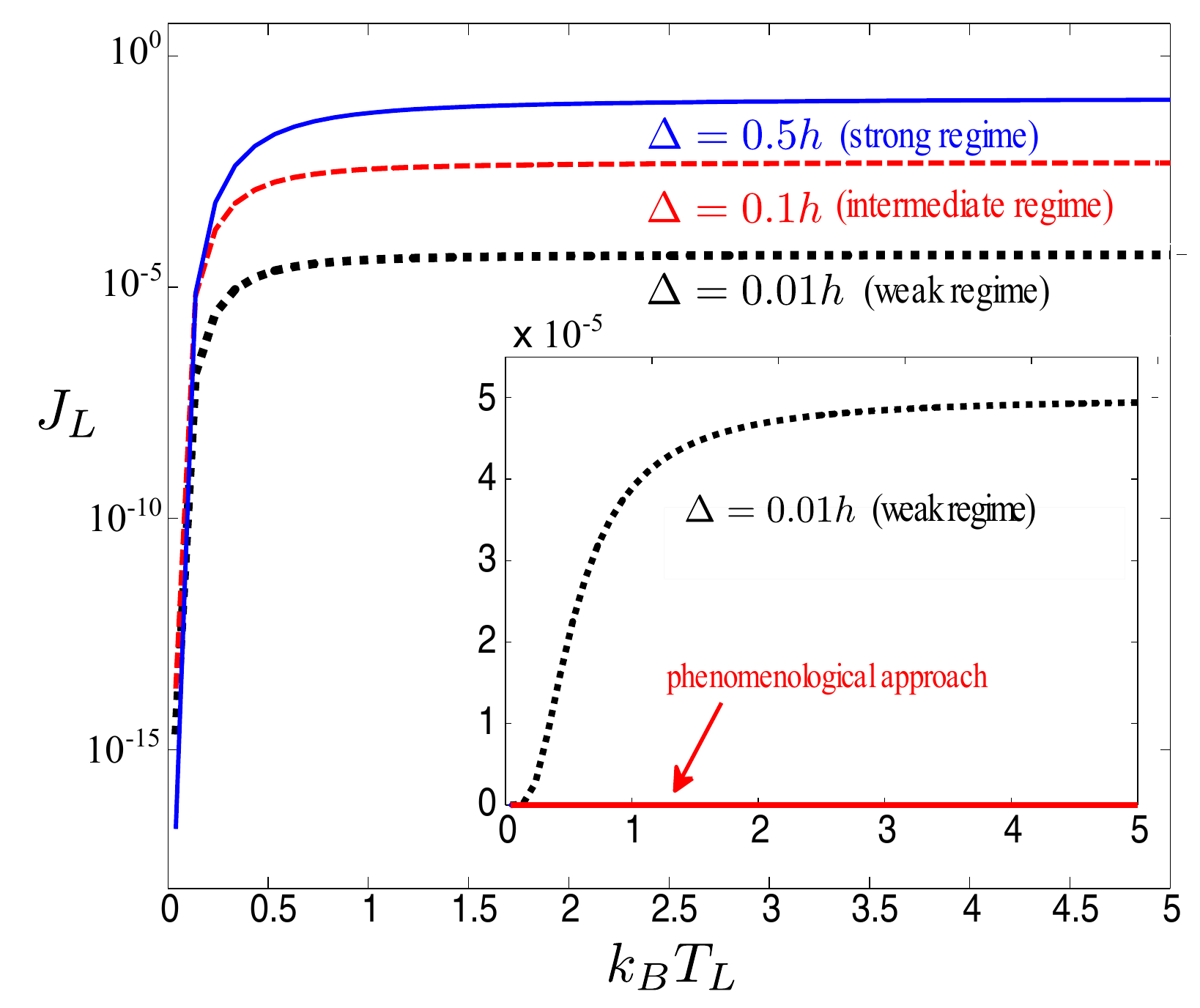}
\caption{(Color Online) Left current $J_L$ as a function of left bath temperature $k_BT_L$ for $T_R=0$ and $\Delta=0.01h$ (black-dotted line, weak regime), $0.1h$ (red-dashed line, intermediate regime), and $0.5h$ (blue-solid line, strong regime). \textit{Inset:} Left current $J_L$ from the microscopic approach (black-dotted line) and the phenomenological approach (red-solid line) as a function of $k_BT_L$ for $T_R=0$ in the weak coupling regime.} 
\label{fig2}
\end{center}
\end{figure}

%
%
{\it Relevance of the strong-coupling formalism.}
We evidence the need for the use of strong-coupling formalism as far as an appropriate description of heat flow in a spin chain is concerned.
In Fig.\ref{fig2}, $J_L$ is shown as a function of $T_L$, at $T_R=0$. We compare three regimes, namely, the strong-coupling regime $\Delta = 0.5h$ (blue, solid line), the moderate or intermediate regime $\Delta=0.1h$ (red dashed line) and the low-coupling or weak regime $\Delta = 0.01 h$ (black, dotted line). For all regimes, the current increases until saturating at a stationary value $J_L = \kappa \Delta^2/2$, since $\Gamma = - \kappa \Delta/4$ for $T_L \rightarrow \infty$ and $T_R \rightarrow 0$ \cite{gammaexpression}. Saturation in the energy flux is expected since the system has a finite number of energy levels. 

The phenomenological model is obtained by replacing the Lindbladians operators in Eq. (\ref{master}) by
\begin{eqnarray*}
\mathcal{L}_{L}^{ph}[\rho] &=& \mathcal{J}(h) (1+n^{L}_h) 
\bigg[ \sigma_{L}^-\rho \sigma_{L}^- -\frac{1}{2}\Big\{ \rho,\sigma_{L}^+  \sigma_{L}^- \Big\} \bigg] \nn \\
                                     & & + \mathcal{J}(h) \  n^{L}_h
                                     \bigg[\sigma_{L}^+ \rho \sigma_{L}^- - \frac{1}{2}\Big\{\rho,\sigma_{L}^- \sigma_{L}^+\Big\} \bigg].
\end{eqnarray*}
and $\mathcal{L}^{ph}_{R}[\rho]$, which is equal to the $\mathcal{L}^{ph}_{L}[\rho]$ with $R$ instead of $L$ and $h\rightarrow0$. This model counterintuitively predicts zero heat current \cite{sup}, $J_L^{ph} = 0$, regardless of the temperature gradient, $T_L-T_R$, and of the coupling constant, $\Delta$. In the inset of Fig.\ref{fig2}, a comparison of the two approaches is made, particularly in the weak-coupling limit, $\Delta = 0.01h$ (red solid line for phenomenological, black dotted line for strong-coupling model).

In order to situate our results with respect to the pertinent literature, 
we analyze the behavior of heat current within another type of spin coupling. In Ref.\cite{briegel}, for instance,
a XY model in a transverse field \cite{modelXY} is considered, for a system of four spins, in which the coupling rates are of the same order of magnitude as the frequencies of the isolated spins, $\Delta \sim h$. However, the heat baths are modeled by the phenomenological method. Fig.3 of Ref.\cite{briegel} indicates that heat current firstly increases to a maximal value and then vanishes, with respect to the gradient of temperature (varying $T_L$, keeping $T_R=0$). That is, the higher the gradient, the smaller the current, for high $T_L$. We show in \cite{sup} that such unexpected behavior is due to the use of a phenomenological approach outside its range of validity. It is worth to highlight that our formalism solves the apparent paradox in the XY model. It yields a physically sound prediction, where the current saturates at a finite value, proportional to the number of quantum levels of the system, not decreasing with respect to the increase of temperature gradient \cite{sup}.

%
%
{\it Optimal rectification.}
Finally, asymmetric conduction of heat as a function of the temperature gradient is studied. 
We focus back on the system formed by two interacting spins as described via Ising model.
We start in the most asymmetric scenario, where $T_L \rightarrow \infty$ and $T_R \rightarrow 0$.
Then, the cycle rate simplifies to $\Gamma = - \kappa \Delta/4$ \cite{gammaexpression}, and the current to $J_L = \kappa \Delta^2/2>0$.
Therefore, finite conduction is established from the left to the right.
On the opposite limit, $T_L \rightarrow 0$, we find that $\Gamma \rightarrow 0$, irrespective of $T_R$. 
Hence, perfect thermal isolation takes place, $J_L(T_L\rightarrow 0) = 0$, when the temperature gradient is reversed. 
Optimal rectification is, therefore, present in that scenario. Interestingly, the phenomenological method predicts vanishing current in both directions \cite{sup}, hence null rectification. This reinforces the disparity between the formalisms.

We give now a clear picture of the asymmetry found above.
The key point reside in transition $\ket{2} \rightarrow \ket{3}$, as it involves heat exchange in the opposite sense with respect to the net heat flow.
Take, for instance, the case $T_L\rightarrow 0$ (cold reservoir) and $T_R>0$ (hot reservoir), as illustrated in Fig.\ref{fig:scheme}(c). 
The natural path for the flow of heat is from the hot reservoir to the cold one. 
In order to accomplish that, heat must jump from the hot reservoir, placed on the right, into the system by flipping the right spin. 
In principle, two paths allow this jumping. 
One is the excitation from state $\ket{1}$ to $\ket{2}$ and the other is from $\ket{3}$ to $\ket{4}$. 
Both steps depend not only on the supply of thermal energy by the reservoir on the right, but also on the population of the  state of departure. 
State $\ket{1}$ is the ground state, so it always have finite probability of being populated. 
Nevertheless, the population of $\ket{3}$ critically depends on the existence of an excitation process from state $\ket{2}$ to $\ket{3}$.
We remind that $\ket{2}=\ket{\dn \dn}$ and $\ket{3} = \ket{\up \dn}$. 
Thus, to pass from state $\ket{2}$ to $\ket{3}$, the system needs to gain energy by flipping the spin on the left. 
Only the left bath flips the left spin.
If the left bath is cold, $T_L\rightarrow 0$, it is not able to provide the required energy for the left spin to flip.
As a result, heat flow from the hot bath on the right to the cold one on the left is blocked precisely at transition $\ket{2}\rightarrow \ket{3}$ (formally, $\Gamma^L_{23} = \Gamma = 0$).

If the gradient is reversed so the left bath is hot, $T_L>0$, and the right bath is cold $T_R\rightarrow 0$,  the natural path for the flow of heat also gets reversed, as illustrated in Fig.\ref{fig:scheme}(b).
In order to allow a finite flow of heat in the new direction, the system has to gain excitation from the left, $\ket{1}=\ket{\dn \up}\rightarrow\ket{4}=\ket{\up \up}$ and lose excitation to the right bath. 
It must then execute the decays $\ket{4}\rightarrow \ket{3}$ and $\ket{2} \rightarrow \ket{1}$.
Again, this is only possible if an intermediate transition occurs, i.e., $\ket{3}\rightarrow \ket{2}$.
This involves loosing energy to the hot reservoir on the left. But that happens with finite probability per unit of time ($\Gamma^L_{32} = -\Gamma = \kappa \Delta/4>0$), in contrast to the absorption of heat from a cold reservoir as in the previous case.

\begin{figure}
\begin{center}
\includegraphics[width=1.00\linewidth]{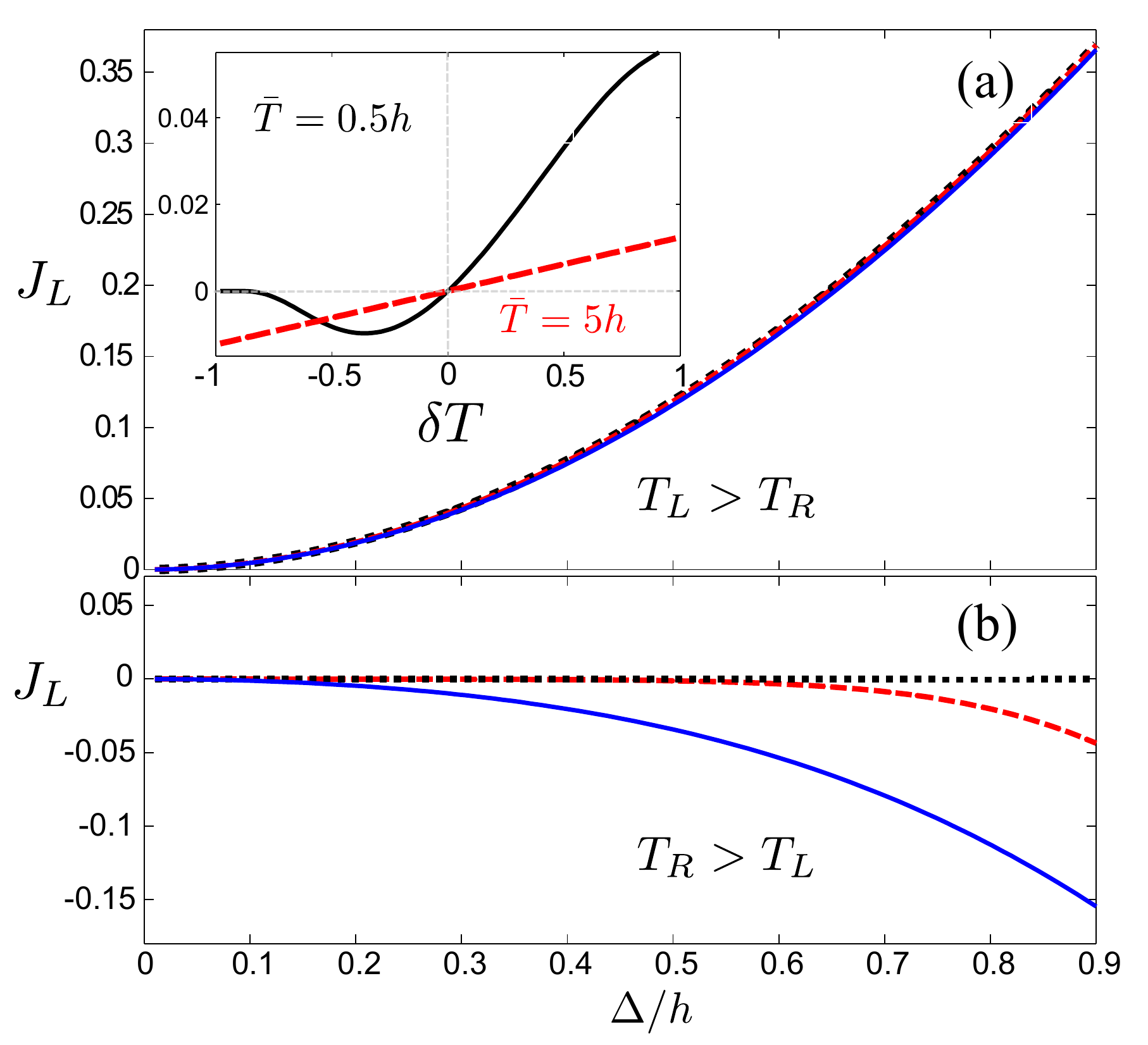}
\caption{(Color Online)(a) Left current $J_L$ as a function of $\Delta/h$ for (a) $k_BT_L=10h$ and $k_BT_R=0,0.1h,0.3h$, and (b) $k_BT_R=10h$ and $k_BT_L=0,0.1h,0.3h$. We used a black-dotted line for $k_BT_{R(L)}=0$, a red-dashed line for $k_BT_{R(L)}=0.1h$, and a blue-solid line for $k_BT_{R(L)}=0.3h$. When the temperature of the cold reservoir is set to absolute zero, we find optimal rectification for any value of the coupling parameter $\Delta$. On the other hand, if the temperature of the cold reservoir is finite, it is necessary to reduce the coupling parameter to achieve optimal rectification. \textit{Inset:} Left current $J_L$ as a function of the temperature gradient $\delta T=k_B(T_L-T_R)$ for $\Delta=0.5h$ and $\bar{T}=0.5h$ (black-solid line), and $5h$ (red-dashed line). Here $\bar{T}= k_B(T_L+T_R)/2$ denote the average temperature. As the average temperature decreases, the curve around $\delta T=0$ becomes asymmetric, indicating the presence of rectification.}
\label{fig3}
\end{center}
\end{figure}

Fig.\ref{fig3} proves that optimal rectification is robust to the more realistic scenario where the cold reservoir is not exactly at zero temperature.
We compare $J_L(\Delta)$ in the cases $k_B T_L=10h \gg k_B T_R = 0h, 0.1h, \mbox{and} \ 0.3h$, in Fig.\ref{fig3}(a), resp. black dotted, red dashed and solid blue curves, and $T_L= 0h, 0.1h, \mbox{and} \ 0.3h\ll T_R=10h$, in Fig.\ref{fig3}(b), resp. black dotted, red dashed and solid blue curves. 
Perfect isolation of heat flow still holds at $k_B T_L \sim 0.1h$ ($k_B T_R = 10h$), for couplings below $\Delta \sim 0.5 h$. Hence, optimal rectification can be preserved even if both reservoirs have nonzero temperature. If the temperature of the cold bath raises above $T = 0.3h/k_B$, asymmetric conduction is suddenly reduced. We can understand this result based on the previous discussion. The existence of a heat flux when $T_R> T_L$ depends on the transition $\ket{2} \rightarrow \ket{3}$. However, as the energy associated with this transition is $\omega_{32}=h-\Delta$, the left current $J_L$ will be approximately zero if the thermal energy supplied by the left reservoir, $k_BT_L$, is much smaller than $\omega_{32}$.

The inset of Fig.\ref{fig3} illustrates the characteristic curve of the thermal diode, $J_L(\delta T)$ at constant  $\bar{T}\equiv k_B (T_L+T_R)/2$, where $\delta T \equiv k_B (T_L-T_R)$. 
Two average temperatures are considered: $\bar{T}_\mathrm{low} = 0.5h$ (black solid curve) and $\bar{T}_\mathrm{high}=5h$ (red dashed). 
Asymmetry is evidently guaranteed at low average temperatures, whereas it completely ceases at high average temperatures.

%

%
{\it Conclusions.}
We derive a master equation for a two-spin chain, within the Ising model, valid in the strong-coupling regime 
$\Delta \sim h$.
This formalism proves necessary for correctly describing heat flow, yielding  $J_L(\delta T)>0$ at $\delta T>0$ not only in the strong-coupling regime $\Delta \sim h$, but also in the weak-coupling limit, $\Delta \ll h$.
Optimal rectification of heat current is predicted in the limit $T_{L(R)}\rightarrow 0$ with $T_{R(L)}\rightarrow \infty$.
An intuitive explanation of that asymmetry is given.
Optimal rectification is robust with respect to nonzero low temperatures ($T\sim 0.1h/k_B$) of the cold thermal bath.
The phenomenological formalism predicts no heat flow in any direction, for the same parameter values to which rectification is optimal in the strong-coupling approach.
Application of this effect to practical devices should allow, for example, temperature control of quantum circuits.
Among the perspectives offered by this work are the influence of strong-coupling in the quantum Fourier law and in quantum thermal machines.

%
{\it Acknowledgements.}
DV thanks E. Mascarenhas for fruitful discussions. 
TW, MC and DV acknowledges financial support from CNPq, Brazil. 
MC acknowledges financial support from INCT-Informa\c c\~ao Qu\^antica, Brazil.

%
%


\begin{thebibliography}{10}
\bibitem{modelXY} This model is described by the Hamiltonian $H_{xy}=\Delta\sum_{i=1}^{N-1}\left(\sigma_x^i\sigma_x^{i+1}+\sigma_y^i\sigma_y^{i+1}\right)/2 +h\sum_{i=1}^N\sigma^i_z/2$, where $N$ denotes the number of spins.
\end{thebibliography}
\end{document}